\documentclass[%
 reprint,
 amsmath,amssymb,
 aps,
]{revtex4-1}

\usepackage{bm}
\usepackage{graphicx}%
\usepackage{color}%

\usepackage{bm}
\usepackage{graphicx}%
\usepackage{color}%

\begin{document}


\title{The Hubbard Model on a Triangular Lattice: \\
Unconventional superconductivity with a nearly flat band}%

\author{Huaiming Guo$^{1}$}
\author{Xingchuan Zhu$^{2}$}
\author{Shiping Feng$^{2}$}
\author{Richard T. Scalettar$^{3}$}

\affiliation{$^1$Department of Physics, Key Laboratory of Micro-Nano
Measurement-Manipulation and Physics (Ministry of Education), Beihang
University, Beijing, 100191, China}

\affiliation{$^2$Department of Physics, Beijing Normal University,
Beijing, 100875, China}

\affiliation{$^3$Physics Department, University of California, Davis,
Ca 95616, USA}



\begin{abstract}
The pairing symmetry of the Hubbard Hamiltonian on a triangle lattice
with a nearly-flat low energy band is studied with the determinant
quantum Monte Carlo method.  We show that the low temperature phase is
insulating at half-filling, even for relatively weak interactions.  The
natures of the spin and pairing correlations upon doping are determined,
and exhibit an electron-hole asymmetry.  Among the pairing symmetries
allowed, we demonstrate that the dominating channels are $d$-wave,
opening the possibility of condensation into an unconventional
$d_{x^2-y^2}+id_{xy}$ phase, which is characterized by an integer
topological invariant and gapless edge states. The results are closely
related to the correlated insulating phase and unconventional
superconductivity discovered recently in twisted bilayer graphene.
\end{abstract}

\maketitle

\textit{Introduction-}
Over the last decade, studies of bilayer and rotated layer graphene have
revealed a wealth of information concerning the modifications to the
Dirac band structure of a single honeycomb lattice which result from
interlayer hybridization $t_\perp$.
Much of the initial work\cite{bistrizer10,dalaissardiere12,
dalaissardiere10}
explicitly tackled the very large
unit cells associated with small twist angles $\theta$.
Although Bernal (AB) stacked bilayers
lose linear dispersion and
chirality properties, it was shown that these can be restored at
other twist angles.  For intermediate
$2^\circ < \theta < 15^\circ$, for example, Dirac bands with a
renormalized velocity persist.  These calculations helped clarify
experimental observations of graphene-like properties even in
materials with large numbers of
planes\cite{emtsev08,sprinkle09,hicks11,miller09,berger06,sadowski06},
far from the single-layer graphene limit.

Beyond the continued presence of Dirac dispersion, two other
fundamental conclusions were drawn for twisted
graphene bilayers.  First, at certain `magic angles,'
flat bands are formed from the
merger of van Hove singularities on either side of
the Dirac point\cite{dalaissardiere12}.
Second, associated with these flat bands, electronic
states become confined in the
`AA' regions of the Moir\'e pattern formed by the rotation,

Along with these band structure investigations, the effects of
interlayer hybridization on magnetic and superconducting
properties in the presence of an
on-site Hubbard interaction $U$ were
explored\cite{gilbert09,lang12,pujari16,tao14,tang17}.  In a single honeycomb
layer there is a critical value $U_c/t \sim 3.87$\cite{paiva05,otsuka16}
for the onset of antiferromagnetic long range order (AFLRO).  For Bernal
(AB) bilayer stacking, at $t_\perp=t$, it was shown that $U_c/t \sim
2.2$\cite{lang12}, and is accompanied by the opening of a single
particle gap
$\Delta_{\rm sp}$ at a roughly comparable $U/t$.  The presence of sites
with different coordination numbers, $z=3$ and $z=4$, lends an
additional
richness to the magnetic behavior, as does the possibility of quenching
AFLRO through interlayer singlet formation in the (unphysical) regime of
larger $t_\perp$.

These explorations of band structure and magnetism lay an essential
foundation for the very recent discovery of unconventional
superconductivity in magic angle graphene
bilayers\cite{cao18a,cao18b},
which themselves already build on work on novel
pairing in single layers\cite{nandkishore12,baskaran10}.
Indeed, the understanding of the Moir\'e triangular superlattice
of AA and AB sites provides a possible approach to the
understanding of pairing in these systems based on
effective Hamiltonians which treat extended AA and AB regions as `sites'
of a simplified model.

This approach underlies a recent paper which considers topological
superconductivity in a two orbital Hubbard model on
a triangular lattice\cite{xu18}.
Importantly, it opens the door to the use of
Quantum Monte Carlo (QMC) methods,
which can provide an exact
treatment of correlated electron physics, but are limited
to lattices of finite size, and are unfeasible for direct treatment of
the immense unit cells at small $\theta$.
Despite the sign problem\cite{loh90,troyer05,vlad2015},
QMC approaches provided an early, essential clue concerning
$d$-wave pairing in the single band Hubbard
Hamiltonian on a square lattice\cite{scalapino86,white89a,white89b},
and hence, if applicable to an appropriate description of bilayer
graphene, might similarly lend important insight.

In this paper, we apply QMC approaches to
the Hubbard Hamiltonian on a triangle lattice with a
nearly-flat low energy band, which yield results sharing interesting features
with those observed experimentally\cite{cao18a,cao18b}.
Our key conclusions are:
(i)  a correlated insulator arises at half-filling even at relatively
small values of $U/t$;
(ii)  the dominant pairing symmetry is $d$-wave,
degenerate in the $x^2-y^2$ and $xy$ channels, opening the
possibility of a chiral phase;
(iii) (short range) antiferromagnetic fluctuations are present and
on sites participating in the flat band are
significantly stronger below half-filling ($\rho=1$) than above.  Finally,
(iv)  the tendency to superconductivity is also asymmetric,
with a stronger response to doping below half-filling.

In the remainder of this paper we describe our effective model,
providing some additional motivation, discuss its band structure,
and present the qualitative physics within mean field theory, along with
the associated topological properties.  We then turn the results of DQMC
for the Mott gap and magnetic correlations, and, finally,
superconductivity.

\begin{figure}[htbp]
\centering \includegraphics[width=8.5cm]{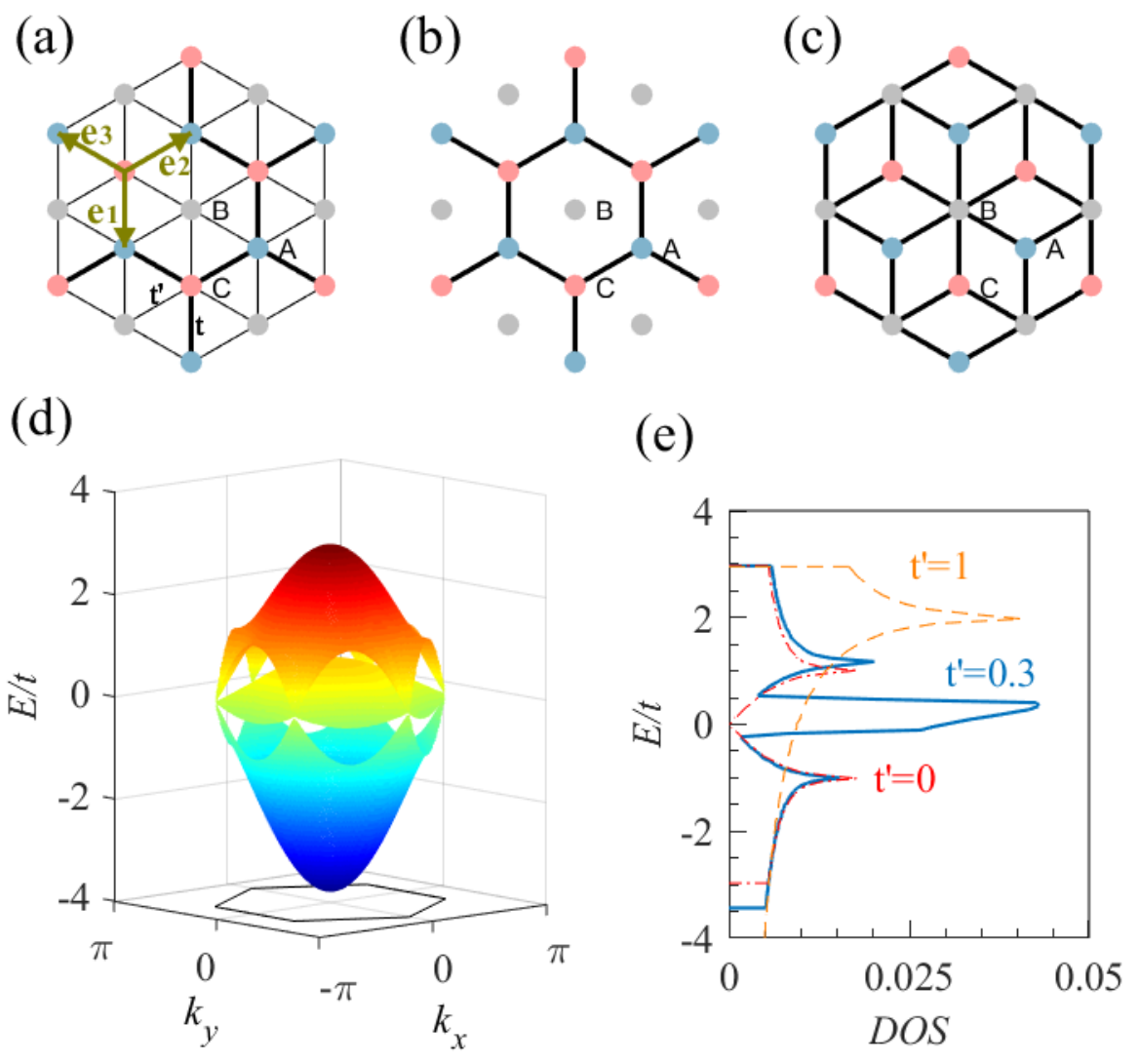} \caption{(a): The effective lattice which results from
treating each AA region as an effective `site' connected by
modulated hopping amplitudes.
Three elementary
vectors ${\bf e}_{1,2,3}$, the $A, B, C$ sublattices are indicated. (b): The honeycomb lattice in the limit $t'=0$. (c): The dice lattice in the limit $t=0$. (d):
The band structure of the effective model in the first Brillouin zone.
(e): The density of states corresponding to the band shown in (d). In (e) we also show the density of states of the limiting cases of the triangular ($t'=t$) lattice and
hexagonal ($t'=0$) lattice. In the latter case,
we do not show the $\delta$-function peak at $E/t=0$.
In (d)
and (e), the anisotropic factor $t'/t=0.3$.
}
\label{fig1}
\end{figure}

\textit{The effective Model-}
Twisted bilayer graphene has been found to have nearly flat low-energy
bands for special discrete angles, where the Moir\'e pattern is a
superlattice comprised of $AA$ and $AB(BA)$ stacking
regions\cite{magic1,magic2,baskaran18,lado18,jiang17}. The wave function is highly concentrated in
the $AA$ regions and is associated with
a band with weak dispersion.  A correlated insulator
is found at half filling\cite{cao18a}.
These considerations suggest the possibility of simulating
each AA region as a `site' in an effective model which includes
a charging energy penalty for occupation of AA regions,
and result in an effective Hubbard Hamiltonian
on a triangle geometry
with modulated hoppings giving rise to a nearly flat low energy band,
\begin{equation}\label{eq1}
H=-\sum_{\langle lj \rangle \sigma}t^{\phantom{\dag}}_{lj}
c^\dag_{j\sigma}c^{\phantom{\dag}}_{l\sigma}+
U\sum_{i}\, \big(n^{\phantom{\dag}}_{i\uparrow}-\frac{1}{2}\big)
\big(n^{\phantom{\dag}}_{i\downarrow}-\frac{1}{2}\big).
\end{equation}
Here $c^\dag_{j\sigma}$ and
$c^{\phantom{\dag}}_{j\sigma}$
are the creation and annihilation operators, respectively, at site $j$
with spin $\sigma=\uparrow, \downarrow$.
$n_{i\sigma}=c^{\dagger}_{i\sigma}c^{\phantom{\dag}}_{i\sigma}$
is the number of
electrons of spin $\sigma$ on site $i$, and $U$ is the on-site repulsion.
The modulation $t,t'$ can be understood by a constuction which begins
with a honeycomb lattice with hopping $t$ and then adding a site at the
center of each hexagon in the honeycomb lattice.  These sites are linked
with hopping $t'$ to their six near neighbors.
See Fig.~\ref{fig1}(a), (b).
Throughout the paper we set to $t=1$ as the unit of energy.

The modified triangle lattice has a three-site unit cell.
In momentum space, the $U=0$ Hamiltonian is,
\begin{equation}\label{eq1K}
{\cal H}_{0}({\bf k})=\left(
                    \begin{array}{ccc}
                      0 & -t\gamma_{\bf k} & -t'\gamma^*_{\bf k} \\
                      -t\gamma^*_{\bf k} & 0 & -t'\gamma_{\bf k} \\
                      -t'\gamma_{\bf k} & -t'\gamma^*_{\bf k} & 0 \\
                    \end{array}
                  \right),
\end{equation}
with $\gamma_{\bf k}=\sum_{j}e^{i{\bf k}\cdot {\bf e}_{j}}$ ($j=1,2,3$).
The spectrum can be directly obtained and contains three branches, as shown
in Fig.~\ref{fig1}(c), (d).  The upper and lower bands have significant
dispersion, while the middle band has a
narrow width,  with a flatness that can be continuously tuned by $t'$.
At either extreme of hopping $t=0$ (dice lattice) and $t'=0$ (honeycomb
lattice), there is a completely flat band intersecting the Dirac points
at zero energy. The flat band in the limit $t'=0$ is formed by isolated
sites, thus is trivial compared to that of the dice lattice.

\begin{figure}[htbp]
\centering \includegraphics[width=8.5cm]{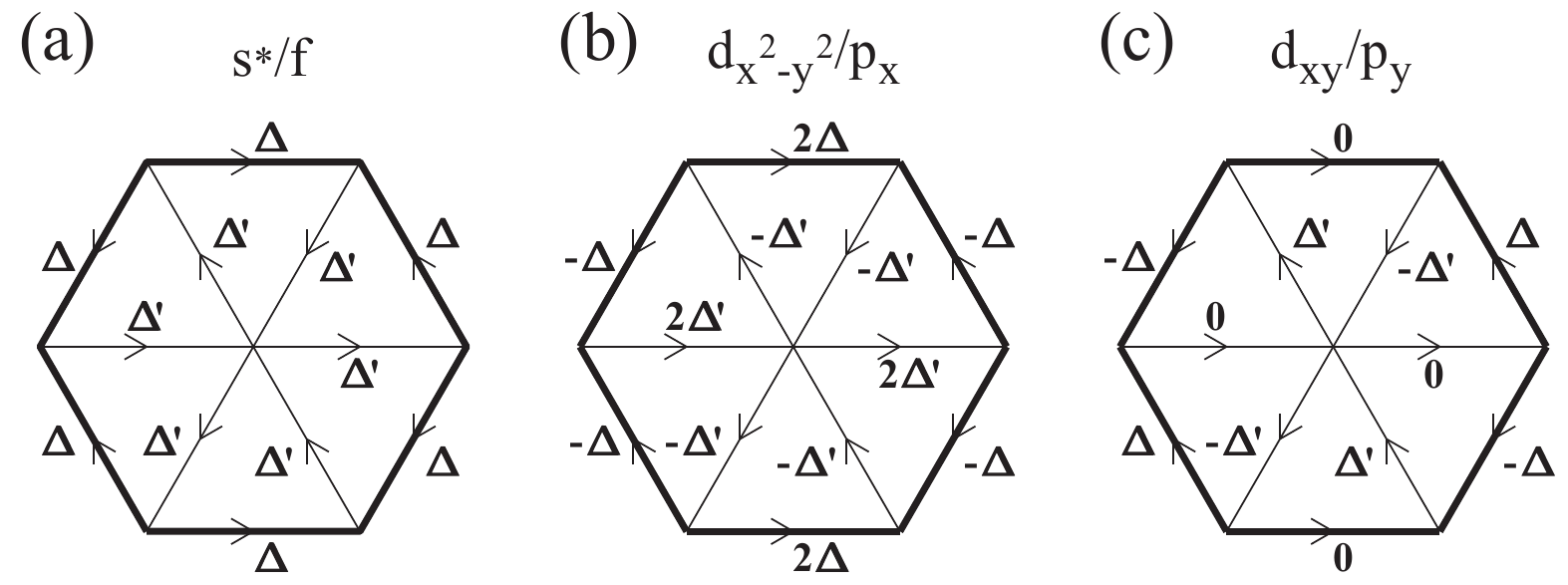} \caption{The pairing
symmetries considered in this paper.  A partner
down spin fermion is created on the
six nearest-neighbor sites of the up spin fermion placed at the hexagon
center. For triplet
pairing, there is an additional sign when the pairing is along the
opposite direction of the arrow.}
\label{fig2}
\end{figure}

\textit{Pairing symmetries and mean-field description of the
superconducting state-}
In the presence of on-site repulsive interactions,
pairing has to be nonlocal.   One can consider a collection of
operators $\Delta^\alpha$ which create an up spin electron
on a site, with a surrounding cloud of down spin electrons on
its near-neighbors.
The pairing symmetries should
be in compatable with the underlying lattice.
The form of the self-consistent BCS gap equation
$\Delta_k = -\sum_{k'} \Gamma_{kk'} (\Delta_{k'}/2E_{k'})
\, {\rm tanh}(E_{k'}/2T)$
for $\Gamma_{kk'}>0$ suggests that only solutions $\Delta_k$ which
change sign (have nodes) in momentum space are allowed\cite{sc2}.
Although the pairing
amplitudes will differ on strong and weak bonds due to the hopping
modulation, the symmetry remains that of the triangle lattice,
i.e.~described by the crystal symmetry group $D_{6h}$ with $k_z=0$. The
possible pairing states can be classified by the irreducible
representations of $D_{6h}$, and include the singlet pairing
symmetries: $s^*$-wave, $d_{x^2-y^2}$-wave, $d_{xy}$-wave, and triplet
pairing symmetries: $p_x$-wave, $p_y$-wave, $f$-wave.  These are
schematically shown in Fig.~\ref{fig2}. Since $d_{x^2-y^2}, d_{xy}$
($p_x, p_y$) belong to the same representation $E_{2g}$ ($E_{1u}$), they
are degenerate, and a linear combination of them is possible when it is
energetically favored.

\begin{figure}[htbp]
\centering \includegraphics[width=8.5cm]{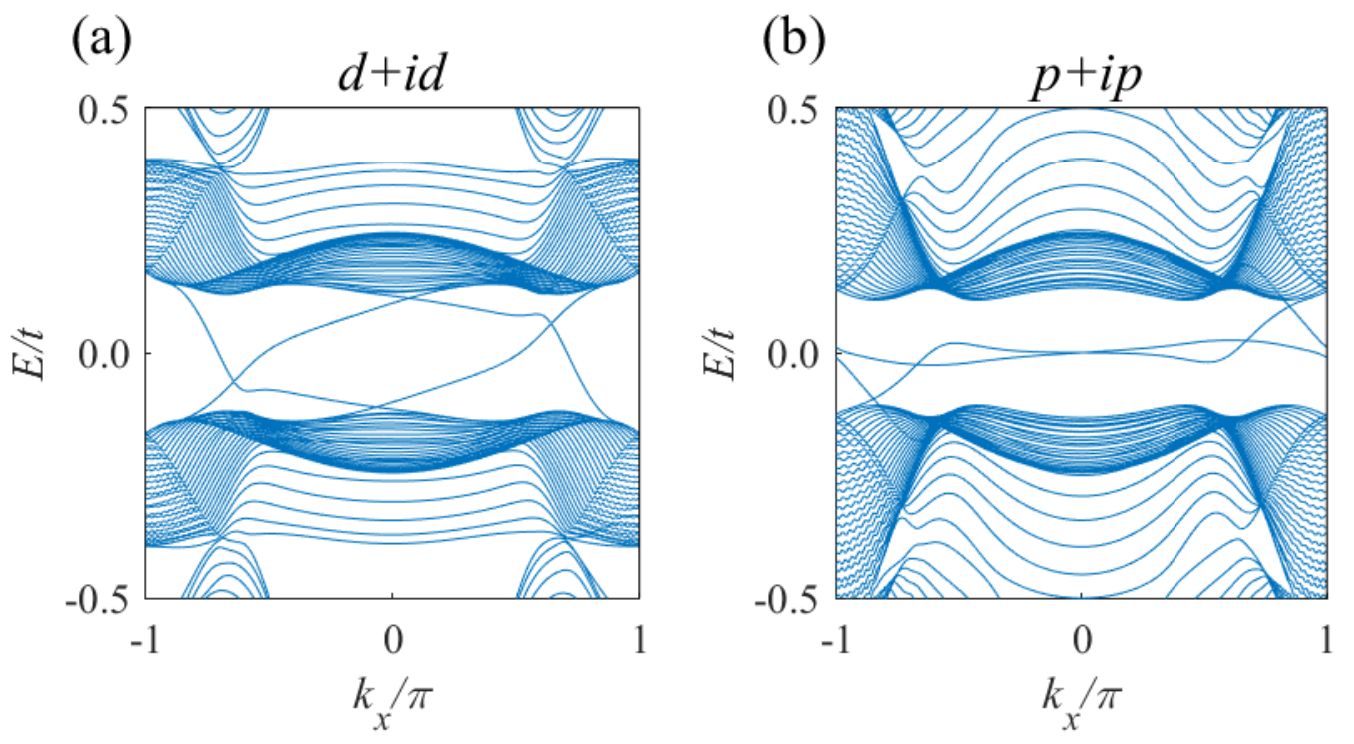} \caption{The
quasiparticle spectrum on zigzag edge ribbon: (a), the $d+id$ chiral
superconducting state; (b), the triplet $p+ip$ state. The parameters are
$t'/t=0.3, \Delta=0.3, \Delta'=0.1$ and $\mu=0.22$ (corresponding to
$\rho=0.94$).}
\label{fig3}
\end{figure}

In the Nambu representation,
the superconducting Hamiltonian in mean-field theory is,
\begin{eqnarray}\label{eq2}
H_{SC}=\sum_{\bf k}\Psi^{\dagger}_{\bf k}{\cal H}_{\bf k}\Psi_{\bf k},
\end{eqnarray}
with $\Psi_{\bf k}=(c_{A,{\bf k}\uparrow}, c_{B,{\bf
k}\uparrow}, c_{C,{\bf k}\uparrow}, c^{\dagger}_{A,-{\bf
k}\downarrow}, c^{\dagger}_{B,-{\bf
k}\downarrow},c^{\dagger}_{C,-{\bf k}\downarrow})^{T}$ and
\begin{eqnarray}\label{eq2L}
{\cal H}_{\bf k}&=&\left(
                   \begin{array}{cc}
                     {\cal H}_{0}({\bf k})-\mu & \Delta^{\dagger}_{\bf k} \\
                     \Delta_{\bf k} & -{\cal H}_{0}({\bf k})+\mu \\
                   \end{array}
                 \right),  \\\nonumber
\Delta_{\bf k}&=&\left(
                   \begin{array}{ccc}
                     0 & \eta_{\bf k} & \zeta\eta_{\bf k}^{'*} \\
                    \zeta \eta^{*}_{\bf k} &0 & \eta'_{\bf k} \\
                      \eta'_{\bf k} & \zeta \eta^{'*}_{\bf k} & 0
                   \end{array}
                 \right).
\end{eqnarray}
Here $\mu$ is the chemical potential.
$\eta_{\bf k}=\sum_{j} \Delta_j
e^{i {\bf k} \cdot {\bf e}_j}$ and $\eta'_{\bf k}=\sum_{j} \Delta'_j
e^{i {\bf k} \cdot {\bf e}_j}$, with pairing amplitudes $\Delta_j$
and $\Delta'_j$ which can be read from the real space arrangement in
Fig.~\ref{fig2}; $\zeta=-1 (+1)$ for singlet (triplet) pairing.  In the
presence of these more complex
interband pairings, the quasiparticle spectrum does not
follow the standard $BCS$ form,
and it is not straightforward to identify
whether there are zero-energy quasiparticles.
By numerically
diagonalizing the Hamiltonian Eq.~(\ref{eq2L}), it is found that the $s^*$-wave
state is fully gapped, and the triplet $f$-wave state has nodes.

Although the $d_{x^2-y^2}$- and $d_{xy}$-wave pairings are gapless, the
chiral one arising from their linear combination
is gapped.
The chiral state is a topological
superconductor characterized by an integer Chern
number\cite{chern},
\begin{eqnarray}\label{eqC}
C=\sum^{occ.}_{n}\frac{1}{2\pi}&&\int_{BZ}dk_xdk_y F_n,  \\ \nonumber
F_n=(\bigtriangledown\times {\bf A}_n)_z&,& {\bf A}_n=i\langle u_{n{\bf k}}|\frac{\partial}{\partial {\bf k}}|u_{n{\bf k}}\rangle.
\end{eqnarray}
Using a gauge-independent method, the Chern number can be directly
calculated numerically\cite{chern1}.  $C=2$ for the
$d_{x^2-y^2}+id_{xy}$ state. In the presence of edges,
gapless states appear which tranverse the gap (see Fig.~\ref{fig3}).
The triplet chiral
$p+ip$-wave state is also topological, with $C=1$.

With this general mean field insight in hand, we turn now to
an explicit evaluation of the superconducting correlation
functions in the different pairing channels.

\textit{DQMC study of the dominating pairing symmetry.-}
The Hubbard model Eq.~(\ref{eq1}) can be solved numerically by means of the
DQMC method\cite{white89a,dqmc}. In this approach, one decouples the on-site
interaction term through the introduction of an auxiliary
Hubbard-Stratonovich field (HSF).  The fermions are integrated
out analytically, and then the integral over the HSF is performed
stochastically.  The
only errors are those associated with the statistical sampling, the
finite spatial lattice
and inverse temperature discretization.
All
are well-controlled in the sense that they can be systematically reduced
as needed, and further eliminated by appropriate extrapolations. The
systems we studied have $N=3\times L \times L$ sites with $L$ up to
$10$.
The sign problem\cite{loh90,troyer05,vlad2015} limits accessible temperatures
unless special symmetries prevent the product of determinants,
which serves as the HSF probability, from becoming negative.

\begin{figure}[htbp]
\centering \includegraphics[height=5.0cm,width=8.7cm]{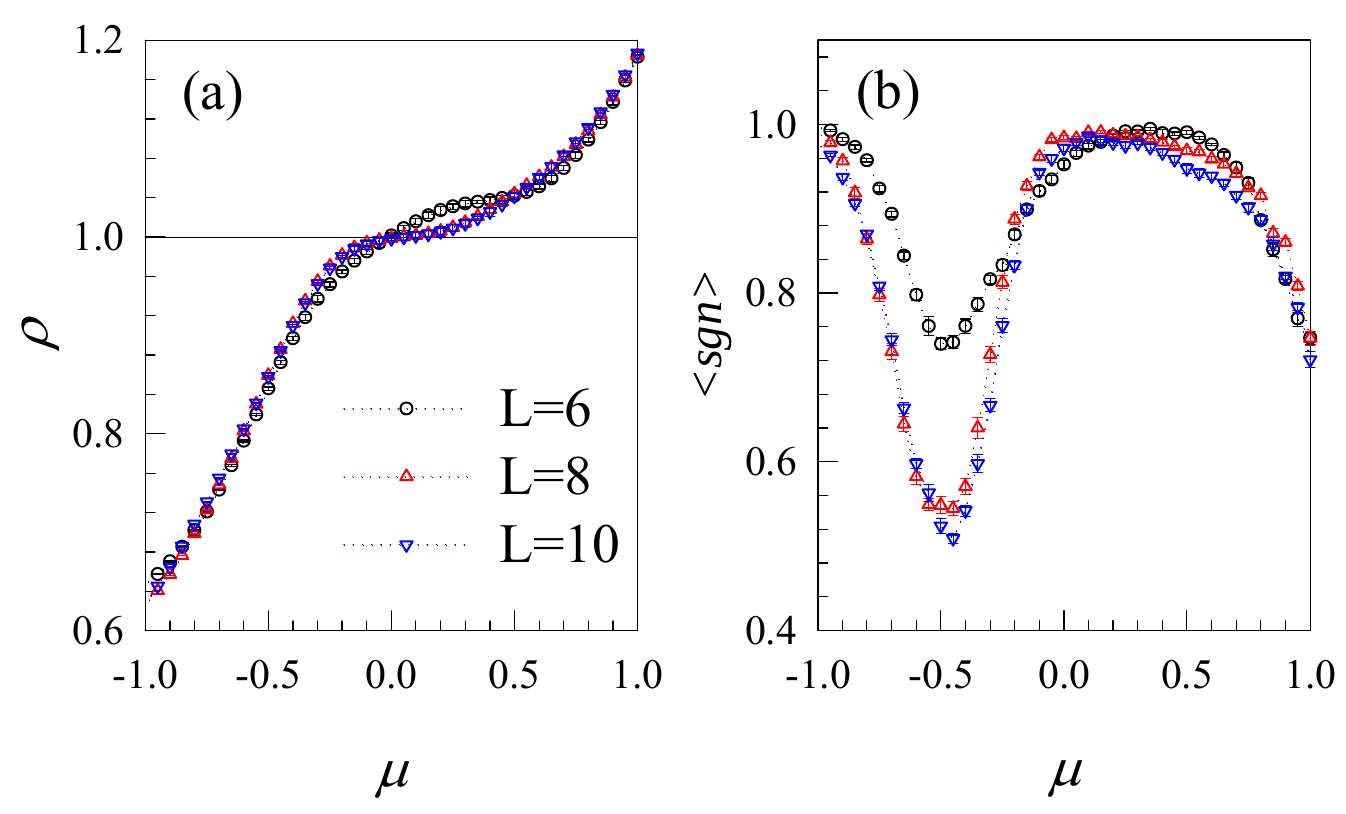}
\caption{
The (a) density and (b) average sign as functions of $\mu$ for linear
lattice sizes $L=6,8,10$. Here $U=2, T=1/12$ and $t'/t=0.3$.
From (a) we see a clear indication of the formation of a
insulating gap at half-filling.  (b) indicates that accessible
temperatures will be limited to $T \gtrsim t/12$ at $\mu \sim -0.5$
}
\label{fig4}
\end{figure}

Figure \ref{fig4} shows the density $\rho$ and average sign
$\langle sgn \rangle$ as functions of
$\mu$ at $U=2$.   At $t'=0$, the geometry consists of a honeycomb lattice
and a collection of independent sites; there is no sign problem.
As $t'$ increases, the lattice is no longer bipartite and
$\langle sgn \rangle < 1$.
As shown in Fig.~\ref{fig4}(b) at $T=1/12, U=2, t'/t=0.3$,
$\langle sgn \rangle \gtrsim 0.6$ over the full range
of densities.  $\rho(\mu)$
has a flat $\rho=1$ region near $\mu=0$ which becomes more
pronounced as the lattice size increases.
This implies that the system exhibits an insulating phase
at half filling, with the gap size $\Delta_{\rm sp} \sim 0.5t$
set by the width of the $\rho=1$ plateau.
(This value is of the same order of magnitude as found in the Bernal case
$\Delta_{\rm sp} \sim 0.3t$ at $U=4$ and $t'=t$.
See Ref.~\cite{lang12}.)
The correlated insulator behavior
is consistent with that recently observed in magic-angle
graphene superlattices \cite{cao18a}, indicating
the model of Eq.~(\ref{eq1}) captures one of the key experimental
features.  A (Slater) gap appears at weak coupling also for a
square lattice.  Its origin there is in the AFLRO
which onsets for any $U>0$ owing to Fermi surface nesting.
For generic geometries without AFLRO, a non-zero $U_c$, set by the bandwidth,
is required to enter the Mott phase.  Here the flatness of the
central band induces strong correlation physics even at small
values of $U$ relative to the total bandwidth.

\begin{figure}[htbp]
\centering \includegraphics[height=5.2cm,width=8.5cm]{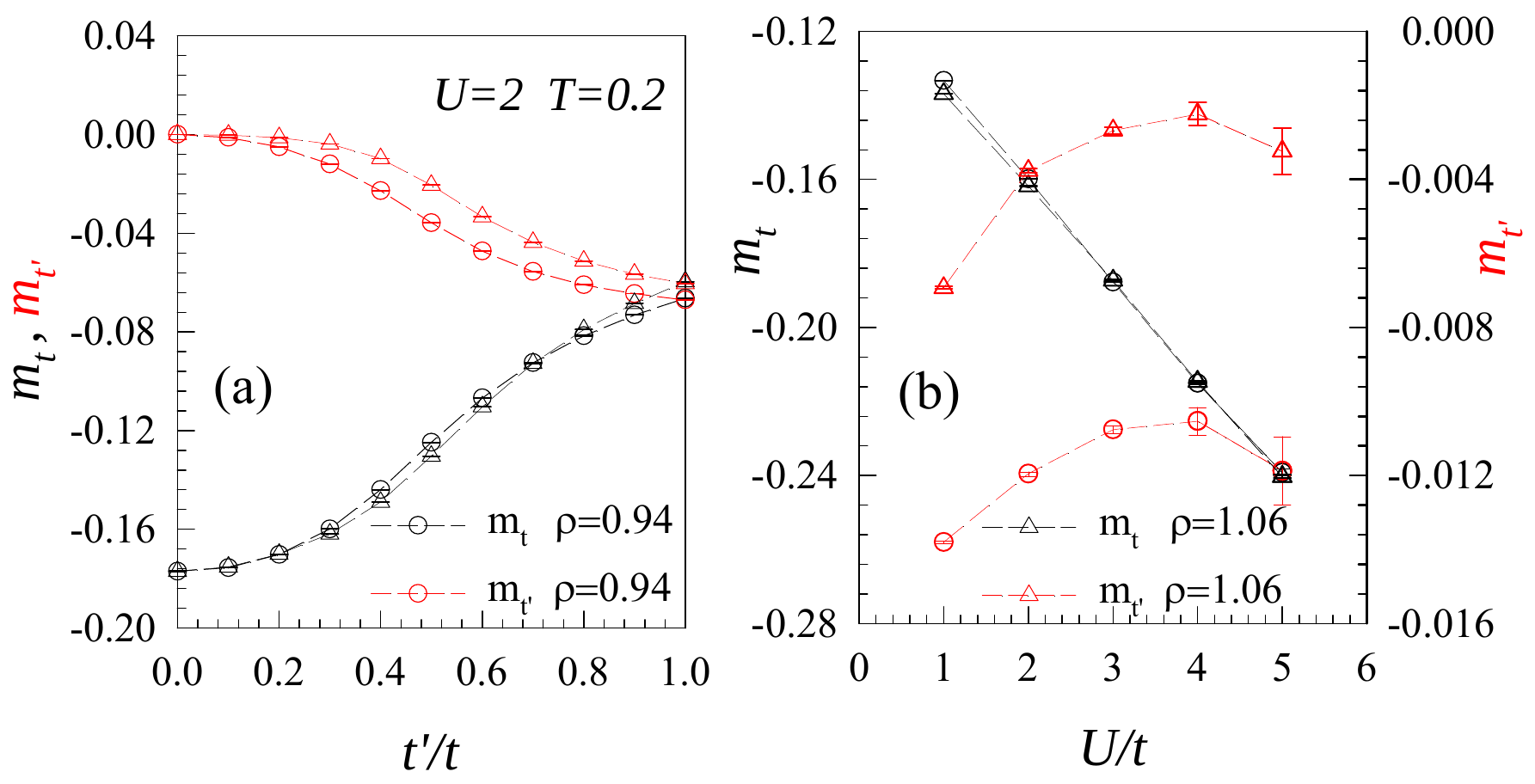} \caption{The
near-neighbor spin correlations $m_{t'}$ and $m_t$
along the $t'$ and $t$ bonds respectively.  (a) as a function of
$t'/t$ for fixed $U=2$.  (b) As a function of $U$ for fixed
$t'/t=0.3$.  The temperature $T=0.2t$.  Results for densities
on either side of half-filling are shown.
}
\label{fig5}
\end{figure}

Short range antiferromagnetic correlations are also present, as seen in
Fig.~\ref{fig5}.
Values are identical
in the triangular lattice limit $t'=t$. $m_t$ grows steadily in
magnitude as $t'$, and hence frustration, are reduced.  $m_{t'}$
decreases with weakening $t'$ [Fig.~\ref{fig5}(a)].
Data for $m_t$ at $\rho=0.94$ and $\rho=1.06$ are virtually
indistinguishable.  However, at $t'/t=0.3$, $m_{t'}$ is roughly
three times larger in magnitude for dopings below $\rho=1$
than for dopings above $\rho=1$ [Fig.~\ref{fig5}(b)].
This suggests a similar asymmetry might occur
for superconductivity which plays off magnetic fluctuations.

To determine the dominating pairing symmetry, we evaluate
the uniform pairing susceptibility,
\begin{eqnarray}\label{eq4}
\chi^{\alpha}=\frac{1}{N}\int_{0}^{\beta}d\tau \sum_{ij}\langle
\Delta^{\alpha}_{i}(\tau)\Delta^{\alpha\dagger}_{j}(0)\rangle
\,,
\end{eqnarray}
The time dependent pairing operator
$\Delta_{i}^{\alpha}(\tau)= \sum_{j}f_{ij}^{\alpha}\,e^{\tau
H}c_{i\uparrow}c_{j\downarrow} e^{-\tau H}$ with $f_{ij}^{\alpha}=0, \pm
1$ or $\pm 2$ for the bond connecting $i$ and $j$, depending on the
pairing symmetry $\alpha$ (Fig.~\ref{fig2}).   The effective
susceptibility $\chi_{\rm eff}^{\alpha}=\chi^{\alpha}-\chi_{0}^{\alpha}$,
subtracts the uncorrelated part $\chi_{0}^{\alpha}$
from $\chi^{\alpha}$, thereby more directly measuring the enhancement
due to $U$.  $\chi^\alpha_{\rm eff}$ can be used to evaluate the
pairing vertex~\cite{sc2,sc3}.

\begin{figure}[htbp]
\centering \includegraphics[height=5.2cm,width=8.5cm]{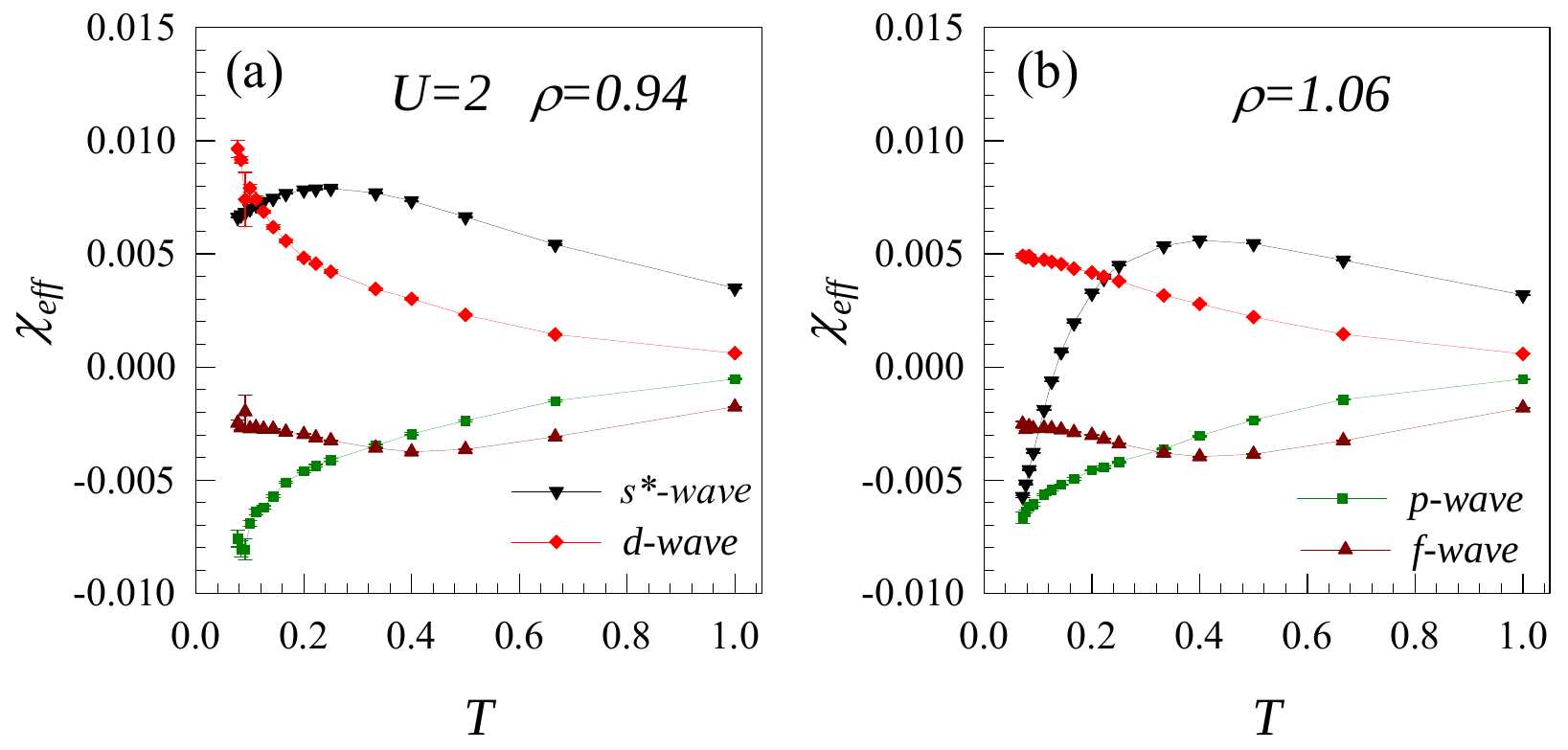} \caption{The
effective pairing susceptibility at $\rho=0.94$ as a function of
temperature for different pairing channels. Here $U=2$ and the lowest
temperature accessed by DQMC is $T=1/13$. Here $\Delta'/\Delta=0.3$ and
the results are similar for other ratios.}
\label{fig6}
\end{figure}

Figure \ref{fig6} shows $\chi^\alpha_{\rm eff}$ vs
temperature for different pairing channels at $\rho=0.94$ and $\rho=1.06$
for $U=2$ and $t'/t=0.3$.
The values for triplet $p\,$- and $f$-wave pairings are negative
(repulsive);
those of the corresponding singlet $s^*$- and $d$-
channels are positive (attractive).
Moreover,
$\chi^d_{\rm eff}$ increases rapidly at low
temperatures (in contrast to the behaviour of
$\chi^{s*}_{\rm eff}$.

$\chi_{\rm eff}$
cannot distinguish
degenerate symmetries, such as $d_{x^2-y^2}, d_{xy}$ and $p_x, p_y$.
Linear combinations will have the same $\chi_{\rm eff}$.
To determine the optimal pairing symmetry, an analysis of the
Ginzburg-Landau free energy such as in Ref.\cite{nandkishore12} should be performed.
From our finite lattice DQMC results, where no spontaneous symmetry
breaking is possible, we can infer only that a
chiral $d_{x^2-y^2}+id_{xy}$ symmetry is a candidate phase.
A qualitative argument in favor of
the chiral phase is that it allows a non-trivial
solution of the gap equation (see discussion above),
while leaving the gap everywhere large.  This suggests
it might be energetically favored\cite{sc2}.

\textit{Conclusions.-}
The appropriate lattice geometry (band structure) and nature of
interactions that need to be incorporated in a Hamiltonian describing
superconductivity in twisted bilayer graphene are, of course, uncertain
at this point.  Suggestions include bilayer triangular and honeycomb
models\cite{xu18,po18,fu18,xyxu18}, and interactions which have SU(4) intra and
inter-orbital symmetry. Studies starting from a
continuum model\cite{bitan18} or considering other pairing
mechanisms\cite{phillips18} have also appeared.
The situation parallels that following the
discovery of cuprate superconductivity, where single band (square
lattice) models contended alongside three band (CuO$_2$) models, and
both on-site $U$ (spin fluctuation) and inter-band $V$ (charge
fluctuation) mechanisms were explored.

In this work,
we have studied the pairing symmetry of a triangular
lattice Hubbard Hamiltonian with modulated hoppings
using the DQMC method.  We first argued
that the band structure of this model incorporates a nearly-flat low
energy band, which underlies the physics of the graphene superlattice,
and then demonstrated that insulating behavior occurs at weak
interactions. Among the pairing symmetries allowed by the triangular
symmetry, the dominating pairing channels are linear combinations of the
degenerate $d_{x^2-y^2}$ and $d_{xy}$ symmetries, including $d_{x^2-y^2}
+ i d_{xy}$ pairing, a form which is topological and characterized by an
integer topological invariant and gapless edge states.

\textit{Acknowledgments.-}
The authors thank W.~Pickett for helpful information. H.G. acknowledges support from NSFC grant No. 11774019. X.Z. and S.F. are supported by the National Key Research and Development Program of China under Grant No. 2016YFA0300304, and NSFC under Grant Nos. 11574032 and 11734002. The work of R.T.S.~is supported by DOE grant No. DE-SC0014671.

\appendix
\section{Evolution of the band structure}

\begin{figure}[htbp]
\centering \includegraphics[width=7cm]{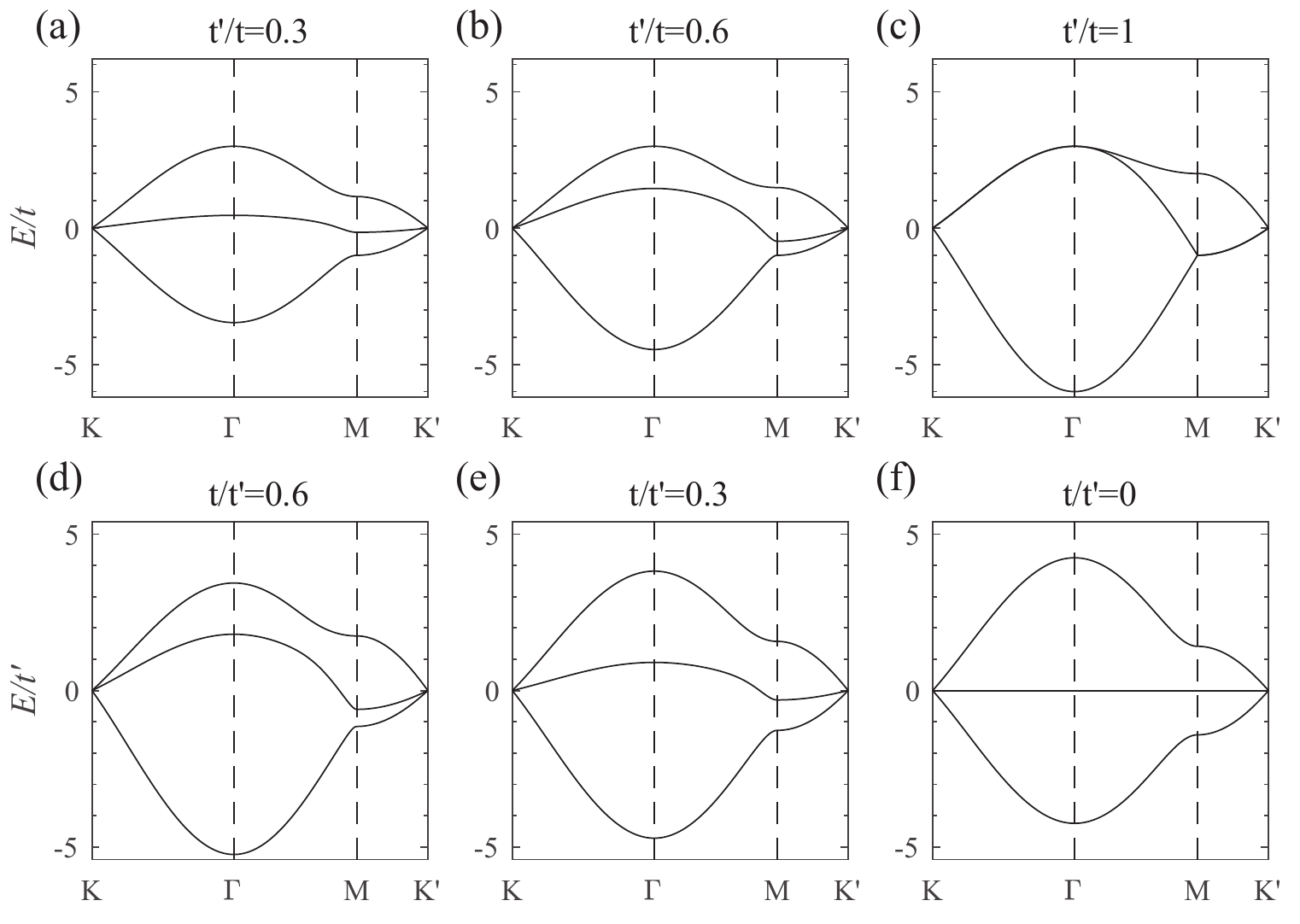} \caption{The evolution of the band structure as a function of the anisotropic ratio $t'/t$.}
\label{fig7}
\end{figure}

The band structure evolves with the anisotropic ratio $t'/t$, which is shown explicitly in Fig.\ref{fig7}. For $t'<t$ the trivial flat band disperses as $t'/t$ increases. The $t=0$ limit is the dice lattice and the Hamiltonian in the momentum space writes as,
\begin{equation}\label{eqa}
{\cal H}_{0}^{dice}({\bf k})=\left(
                    \begin{array}{ccc}
                      0 & 0 & -t'\gamma^*_{\bf k} \\
                      0 & 0 & -t'\gamma_{\bf k} \\
                      -t'\gamma_{\bf k} & -t'\gamma^*_{\bf k} & 0 \\
                    \end{array}
                  \right).
\end{equation}
The energy spectrum contains three branches: $E_{\bf k}^{1(2)}=\pm\sqrt{2}t'|\gamma_{\bf k}|$ and $E_{\bf k}^{3}=0$. The flat band also disperses for $t/t'\neq 0$. At $t=t'$ the band structure becomes that of the triangle lattice.

\section{The superconducting order parameter}
\begin{figure}[htbp]
\centering \includegraphics[width=7cm]{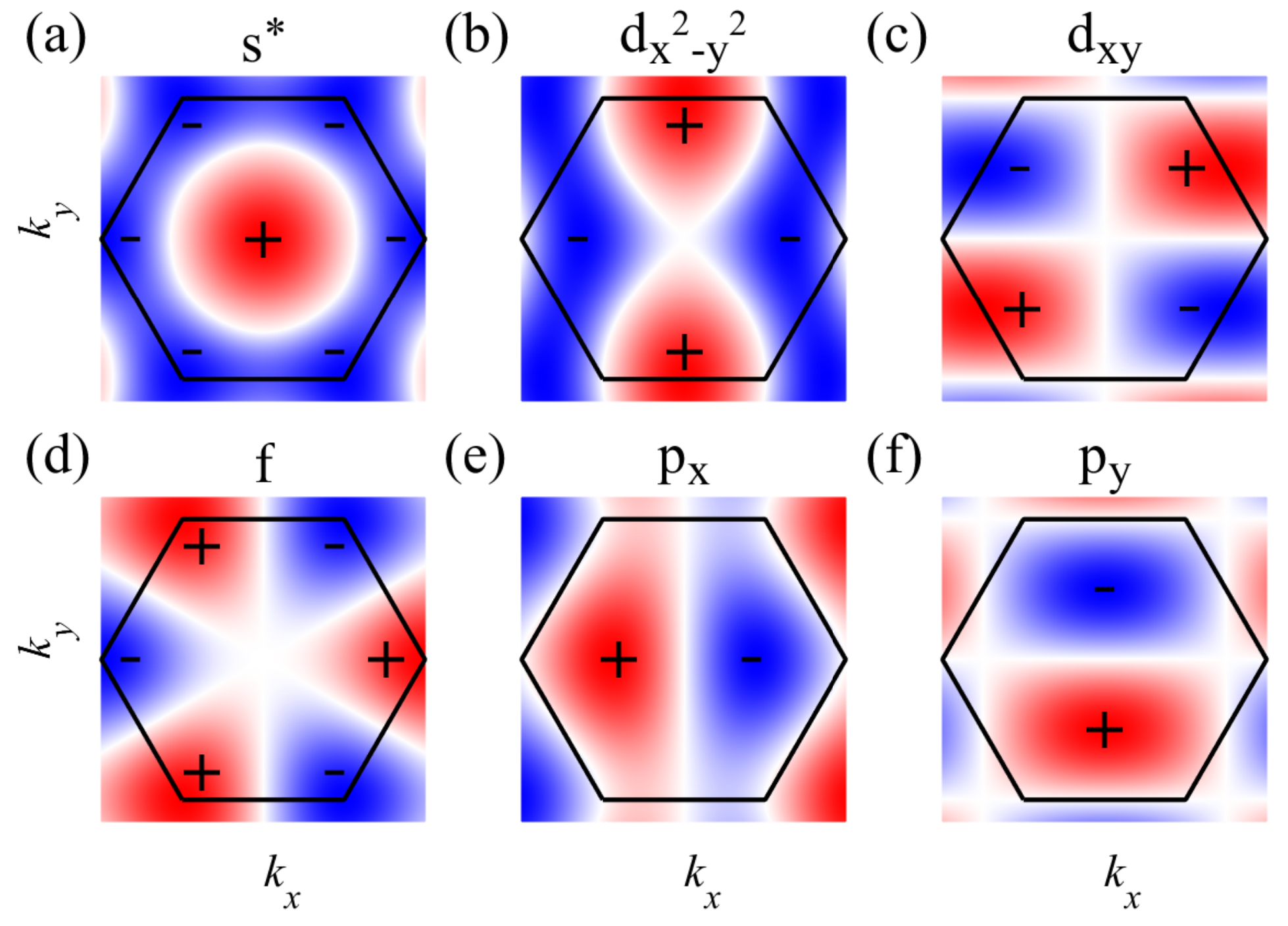} \caption{The momentum dependence of $\Delta_{\bf k}^{t}$ for $s^*$, $d_{x^2-y^2}$, $d_{xy}$, $f$, $p_x$, $p_y$ pairing channels.}
\label{fig8}
\end{figure}

When $t=t'$, the geometry is the normal triangle lattice. The superconducting Hamiltonian in the momentum space is,
\begin{eqnarray}\label{eqb}
{\cal H}_{\bf k}^{t}&=&\left(
                   \begin{array}{cc}
                     {\cal H}_{0}^{t}({\bf k})-\mu & \Delta^{t\dagger}_{\bf k} \\
                     \Delta_{\bf k}^{t} & -{\cal H}_{0}^{t}({\bf k})+\mu \\
                   \end{array}
                 \right),
\end{eqnarray}
Here the noninteracting Hamiltonian is ${\cal H}_{0}^{t}({\bf k})=-t(\gamma_{\bf k}+\gamma^{*}_{\bf k})$. The superconducting order parameter is $\Delta_{\bf k}^{t}=\sum_{j=1}^{3} \Delta_j
(e^{i {\bf k} \cdot {\bf e}_j}+\zeta e^{-i {\bf k} \cdot {\bf e}_j})$ with  pairing amplitudes $\Delta_j$ which can be read from the real space arrangement in Fig.~\ref{fig2}; $\zeta=1 (-1)$ for singlet (triplet) pairing. Figure \ref{fig8} shows the momentum dependence of $\Delta_{\bf k}^{t}$, which is consistent with the symmetries of the corresponding pairing channels.

\begin{figure}[htbp]
\centering \includegraphics[width=7cm]{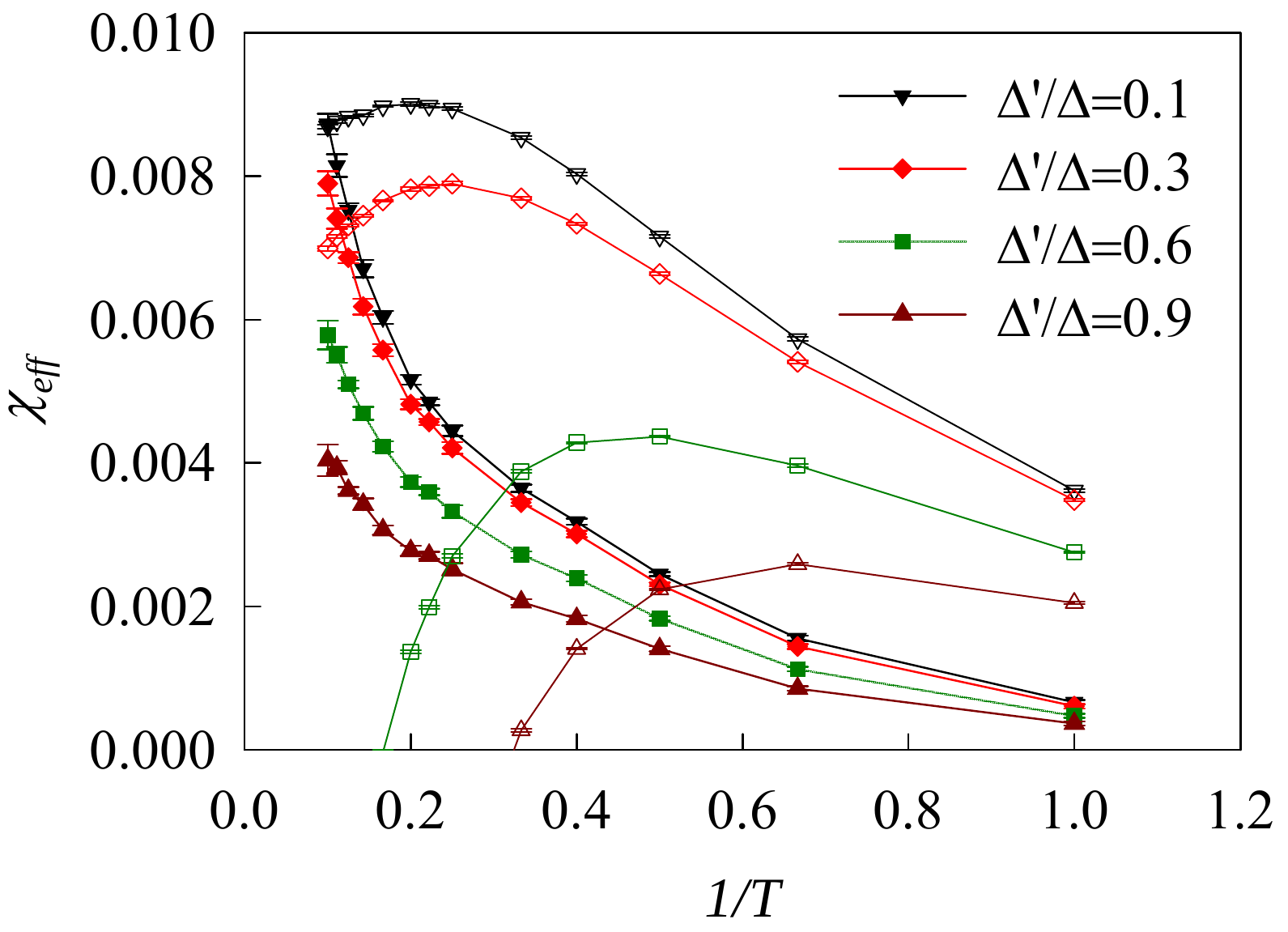} \caption{The effective susceptibility at different $\Delta'/\Delta$ for $s^*$-, $d$-wave pairing channels. The filled (open) symbols represent $d$-($s^*$-)wave pairing phases. The $f$- and $p$-wave phases have smaller $\chi_{eff}$, and are not shown here. The filling is $\rho=0.94$.}
\label{fig9}
\end{figure}

For the case $t'\neq t$, it is expected that the pairing amplitude $\Delta'$ on bonds with $t'$ should be different from that on bonds with the hopping amplitude $t$. However the ratio $\Delta'/\Delta$ can not be determined by our method. We calculate the effective susceptibility for different values of $\Delta'/\Delta$ and find that the $d$-wave phase is always dominate(see Fig.\ref{fig9}).

\bibliography{Biblio}

\end{document}